\documentclass[twocolumn]{aastex631}

\newcommand{\review}[1]{{#1}}

\newcommand{\Rm}{{\rm Rm}}
\newcommand{\Rmeff}{{\rm Rm}_{\rm eff}}

\newcommand{\Ec}{{\rm Ec}}
\newcommand{\En}{E_\eta}
\newcommand{\etat}{\tilde{\eta}}

\usepackage{amsmath}
\usepackage{esint}
\usepackage{ragged2e}
\usepackage{hyperref}
\usepackage{physics}
\usepackage{bm}
\usepackage{nccmath}
\shorttitle{Thermo-resistive instability in hot jupiters}
\shortauthors{Hardy et al.}

\begin{document}

\title{Variability from thermo-resistive instability in the atmospheres of hot jupiters}

\author[0000-0002-2599-6225]{Raphaël Hardy}

\affiliation{Département de Physique, Université de Montréal, Montréal, QC, H3C 3J7, Canada\\}
\affiliation{Department of Physics and McGill Space Institute, McGill University, Montréal, QC, H3A 2T8, Canada\\}
\affiliation{Institut de Recherche sur les Exoplanètes (iREx), Université de Montréal, Montréal, QC H3C 3J7, Canada}

\author[0000-0002-6335-0169]{Andrew Cumming}
\affiliation{Department of Physics and McGill Space Institute, McGill University, Montréal, QC, H3A 2T8, Canada\\}
\affiliation{Institut de Recherche sur les Exoplanètes (iREx), Université de Montréal, Montréal, QC H3C 3J7, Canada}

\author[0000-0003-1618-3924]{Paul Charbonneau}
\affiliation{Département de Physique, Université de Montréal, Montréal, QC, H3C 3J7, Canada\\}

\email{raphael.hardy@umontreal.ca}



\begin{abstract}
The atmosphere of a hot jupiter may be subject to a thermo-resistive instability, in which the increasing electrical conductivity with temperature leads to runaway Ohmic heating. We introduce a simplified model of the local dynamics in the equatorial region of a hot jupiter that incorporates the back reaction on the atmospheric flow as the increasing electrical conductivity leads to flux freezing, which in turn quenches the flow and therefore the Ohmic heating. We demonstrate a new time-dependent solution that emerges for a temperature-dependent electrical conductivity (whereas a temperature-independent conductivity always evolves to a steady-state). The periodic cycle consists of bursts of Alfven oscillations separated by quiescent intervals, with the magnetic Reynolds number alternating between values smaller than and larger than unity, maintaining the oscillation. We investigate the regions of pressure and temperature in which the instability operates. For the typical equatorial accelerations seen in atmospheric models, we find instability at pressures $\sim 0.1$--$1\ {\rm bar}$ and temperatures $\approx 1300$--$1800\ {\rm K}$ for magnetic fields $\sim 10\ {\rm G}$. Unlike previous studies based on a constant wind velocity, we find that the instability is stronger for weaker magnetic fields. Our results add support to the idea that variability should be a feature of magnetized hot jupiter atmospheres, particularly at intermediate temperatures. The temperature-dependence of the electrical conductivity is an important ingredient that should be included in MHD models of hot jupiter atmospheric dynamics.

\end{abstract}



\section{Introduction}\label{sec:intro}

Extreme irradiation from their host stars and tidally locked orbits make for interesting dynamics in hot jupiters' (HJs) atmospheres. The temperature gradient creates a dayside to nightside flow \citep{Komacek2016}, contributing to the driving of equatorial superrotation \citep{Showman2011,Read2018,Imamura2020}. These strong zonal flows displace the hottest part of the surface away from the substellar point. The equatorial superrotation usually leads to eastwards displacement of the hot spot \citep{Showman2002,Cooper2005,Showman2009,Rauscher2010,Kataria2016}. However, in some HJs, westward displacements and variability have been observed \citep{Armstrong2016,Dang2018,Jackson2019,Bell2019,vonEssen2020}.

One explanation for the westward displacements of the hot spot and variability involves magnetohydrodynamical (MHD) effects that reverse the direction of the atmospheric flow \citep{Rogers2014b,Rogers2017b,Hindle2019,Hindle2021a,Hindle2021b}. 
Magnetic effects are expected to be important in HJs atmospheres \citep{Rogers2014b} because the high temperatures lead to partial ionization of the atmosphere, in particular of alkali metals \citep{Batygin2010,Perna2010a}. Magnetic fields have two effects: they can lead to dissipation through Ohmic losses, and they can change the dynamics through $\vec{J}\times\vec{B}$ forces. Simulations have tended to focus on magnetic drag and the associated dissipation, adding a drag term to the momentum equation \citep{Perna2010a} that can significantly slow the atmospheric flow (e.g.~\citealt{Beltz2022}), as well as heating the atmosphere (e.g.~\citealt{Batygin2010}). This is appropriate at low magnetic Reynolds number (lower temperatures), where flux freezing does not hold and Ohmic dissipation dominates advection. \cite{Rogers2014b} included the full dynamics by solving the MHD equations directly, and found that at higher magnetic Reynolds numbers (temperatures), magnetic forces can lead to time-variability, including oscillatory flows, and flow reversals (westward flow). \cite{Rogers2017b} and \cite{Hindle2021a} used these results to derive limits on the magnetic field strength in several HJs.

In this paper, we demonstrate an additional source of variability coming from the temperature-dependence of the electrical conductivity, which leads to a temporal variation of the electrical conductivity as the temperature changes over time. \citet{Menou2012instability} pointed out that the strong increase of the electrical conductivity with temperature allows the development of a thermo-resistive instability (TRI). A small increase in temperature leads to increased electrical conductivity, larger induced currents and therefore enhanced Ohmic heating that further increases the temperature, thus yielding a positive feedback. Describing the flow dynamics with a parametrized magnetic drag law, \citet{Menou2012instability} found local instability at pressures $\approx 3$--$300\ {\rm mbar}$, magnetic field strengths $B\sim 10\ {\rm G}$, and temperatures for which the magnetic drag is weak enough to allow $\approx 10\ {\rm km~s^{-1}}$ zonal wind speeds. Global simulations that include magnetic effects have generally not included a time-dependent electrical conductivity. \cite{Rauscher2013} and \cite{Beltz2022} did include a temperature-dependent magnetic drag force, but none of the studies that solve the full MHD equations included a time-dependent conductivity. 

In this paper, we introduce a simplified model to explore the interplay between the dynamical effect of the magnetic field and the changes in electrical conductivity over time as temperature evolves. We find a dynamical version of the local TRI identified by \citet{Menou2012instability} that leads to time-variability in the flow, either in the form of Alfvén oscillations or short bursts separated by longer quiescent intervals. The key ingredient is the temperature-dependent conductivity, since it allows the magnetic Reynolds number to cycle between small and large values during the oscillation. We then discuss the application of our results to HJ atmospheric conditions and the implications for HJ atmospheric dynamics.

\section{Local model with a time-dependent magnetic diffusivity}
\label{sec:model}

To model the local dynamics in the equatorial plane, we consider the evolution of a zonal flow in the presence of a background radial magnetic field $B_r$. Rather than assume a fixed velocity for the zonal flow, since we are interested in the dynamics, we assume that it is forced by a net flow of momentum into the equatorial region, represented by an acceleration $\dot{v}$. This  could correspond to either an eastward or westward acceleration depending on the overall global dynamics \citep{Showman2011,Hindle2021b}.

\review{Working under the MHD approximation \citep{Davidson2001}, we solve the zonal components of the axisymmetric MHD equations in the equatorial plane of the planet. Thus, all radial and latitudinal flow displacements are eliminated from the system. For the background magnetic field, we consider a purely radial magnetic field\footnote{\review{For the purposes of our simplified model, we consider a background radial field only. A global aligned dipole field would have only horizontal components at the equator, but radial field components would be expected if the dipole is misaligned, or higher order field components are present.}} respecting ${\bf\nabla}\cdot{\bf B}=0$. In this geometry, radial shear drives torsional Alfvén oscillations.}

\review{As the thickness of the atmospheres of HJs are very small compared to their radii, we can use the plane parallel approximation to write our first set of equations. With these approximations in mind, the MHD equations involving the local values of the zonal flow of the fluid in the reference frame of the rotating planet, $U_\phi$, toroidal magnetic field, $B_\phi$, and temperature, $T$, are:}
\begin{equation}\label{eq:movement_full}
\frac{\partial U_\phi}{\partial t} =\frac{B_r}{4 \pi \rho} \frac{\partial B_{\phi}}{\partial r} + \dot{v},
\end{equation}
%
\begin{equation}\label{eq:induction_full}
\frac{\partial B_{\phi}}{\partial t} =  B_r \frac{\partial U_\phi}{\partial r} +\eta  \frac{\partial^2 B_\phi}{dr^2},
\end{equation}
%
\begin{equation}\label{eq:temperature_full}
\rho c_p\frac{\partial T}{\partial t} =  -\rho c_p \frac{(T-T_0)}{\tau} +  \frac{\eta}{4 \pi} \left(  \frac{\partial B_{\phi}}{\partial r} \right) ^2,
\end{equation}
\review{where $\eta$ is the magnetic diffusivity, $\rho$ the gas density and $c_P$ the heat capacity. In the heat equation (\ref{eq:temperature_full}), the first term on the right-hand side represents Newtonian cooling to a reference temperature $T_0$ on a timescale $\tau$, and the second term is Ohmic heating. We take the constant acceleration $\dot{v}$ to be positive (eastward), but our results apply equally to the westward $\dot{v}<0$ case with the direction of flow reversed.}

We now assume that radial variations are on the scale of the thickness of the layer $H$, writing the radial derivative of a quantity $f$ as $\partial f / \partial r \sim f / H$. We can then write down a simplified version of the MHD equations:
\begin{equation}\label{eq:momentum_dim}
	\frac{d U_\phi}{dt} = -\frac{B_r}{4\pi \rho}\frac{B_\phi}{H} + \dot{v} = - \frac{v_A^2}{H}\frac{B_\phi}{B_r} + \dot{v},
	\end{equation}
	\begin{equation}\label{eq:induction_dim}
	\frac{dB_\phi}{dt} = B_r \frac{U_\phi}{H} - \eta \frac{B_\phi}{H^2},
	\end{equation}
	\begin{equation}\label{eq:temperature_dim}
	\rho c_p\frac{dT}{dt} = -\rho c_p\frac{(T-T_0)}{\tau} + \frac{\eta}{4 \pi}\left(\frac{B_\phi}{H}\right)^2,
	\end{equation}
where the Alfvén speed $v_A$ is given by $v_A^2=B_r^2/4\pi\rho$.

To capture the temperature-dependence of the magnetic diffusivity $\eta$, we write 
\begin{equation}\label{eq:diffusivity}
\eta \left( T \right) = \eta_0\exp\left[ E_\eta \left(\frac{T_0}{T} -1\right)\right],
\end{equation}
where $\eta_0 = \eta(T_0)$ and the parameter $E_\eta$ controls the temperature sensitivity ($d\ln\eta/d\ln T = -E_\eta T_0/T$). The exponential form for $\eta(T)$ is similar to the diffusivity from alkali metal ionization, which involves the thermal ionization fraction (e.g.~\citealt{Menou2012a}).

We write the evolution equations in dimensionless form by defining $v=U_\phi/v_A$, $b = B_\phi/B_r$, $\dot{V} = \dot{v} H^2/\eta_0 v_A$, the magnetic Reynolds number $\Rm=v_A H/\eta_0$, and the Eckert number $\Ec=v_A^2/c_p T_0$. We measure time in units of the Ohmic diffusion time $H^2/\eta_0$, and temperature with respect to the reference temperature $T_0$, setting $\eta_0 t/H^2\rightarrow t$, $\eta_0 \tau / H^2 \rightarrow \tau$, and $T/T_0\rightarrow T$. In these units, the Alfvén timescale $H/v_A$ is $1/\Rm$. If we estimate the cooling timescale as $H^2/\kappa_T$, with $\kappa_T$ the thermal diffusivity, then the parameter $\tau\approx \eta_0/\kappa_T$. The dimensionless form of equations (\ref{eq:momentum_dim})--(\ref{eq:temperature_dim}) are then:
	\begin{equation}\label{eq:momentum}
	\frac{dv}{dt} = -{\rm Rm} ~b + \dot{V},
	\end{equation}
  	\begin{equation}\label{eq:induction}
	\frac{db}{dt} = {\rm Rm} ~v - \tilde{\eta}{b},
	\end{equation}
	\begin{equation}\label{eq:temperature}
	\frac{dT}{dt} = -\frac{(T-1)}{\tau} + {\rm Ec}~\tilde{\eta} b^2,
	\end{equation}
where $\tilde{\eta}(T)~=~\eta(T)/\eta_0$ is given by equation (\ref{eq:diffusivity}). 
These equations allow us to solve for the evolution of $b$, $v$ and $T$ given the 5 parameters $\Rm$, $\Ec$, $E_\eta$, $\tau$, and $\dot V$.

\review{We solve the nonlinear coupled ordinary differential equation system defined by equations (\ref{eq:diffusivity}) to (\ref{eq:temperature}) as an initial value problem. Towards this end we use a solver from SciPy\footnote{The function is {\tt solve\_ivp} from the {\tt integrate} module. The documentation can be found at {\tt https://docs.scipy.org/doc/scipy/reference/generated/ scipy.integrate.solve\_ivp.html}}, a Python package, using an explicit Runge-Kutta method of order 5.}

In the absence of forcing ($\dot{V}=0$), equations (\ref{eq:momentum}) and (\ref{eq:induction}) describe damped Alfvén waves with (dimensionless) frequency given by $\omega^2 = \Rm^2$ and damping timescale $1/\etat$. Note also that with $\dot{V}=0$, the limit of magnetic drag is recovered by setting $db/dt = 0$, in which case $b=\Rm v/\etat$ and the momentum equation reduces to $dv/dt = -(\Rm/\etat)^2 v$, representing a drag force with associated timescale $(\etat/\Rm)^2$, in agreement with equation~(1) of \cite{Rauscher2013}.   

With forcing included ($\dot{V}>0$), equations (\ref{eq:momentum})--(\ref{eq:temperature}) have a steady state solution
\begin{equation}\label{eq:ss}
    b = \frac{\dot V}{\Rm}; \hspace{1cm} v = \etat~\frac{\dot V}{\Rm^2},
\end{equation}
where $\etat$ is evaluated at the steady-state temperature for which $dT/dt=0$ (i.e. the value of $T$ for which the right hand side of equation (\ref{eq:temperature}) vanishes\review{, which yields $\bm{T-1=}{\rm\bf Ec} \tilde{\bm\eta} \bm{\tau} \bm b^2 $}). The physical balance represented by equation (\ref{eq:ss}) results from the driven zonal flow winding the radial field until the Ohmic dissipation associated with the induced toroidal magnetic component balances the rate of kinetic energy input to the layer.

For a temperature-independent diffusivity, the evolution of the driven system is always to the steady-state given by equation (\ref{eq:ss}), although the nature of the evolution depends on the magnetic Reynolds number. At high $\Rm$ the system moves towards the steady-state via a series of damped Alfvén oscillations, whereas at low $\Rm$ the system evolves smoothly to the steady state. The situation is very different when the magnetic diffusivity depends strongly on temperature, as we discuss in the next section.


\section{Dynamical Thermo-Resistive Instability}\label{sec:results}

We now show that equations (\ref{eq:momentum})--(\ref{eq:temperature}) have unstable time-dependent solutions when the magnetic diffusivity depends strongly on temperature. We first present some representative evolutions and then discuss the regions of the parameter space where the system is unstable.

\subsection{Examples of time-dependent solutions}\label{sec:diff_regimes}

\begin{figure}[t]
    \includegraphics[width=1.0\linewidth]{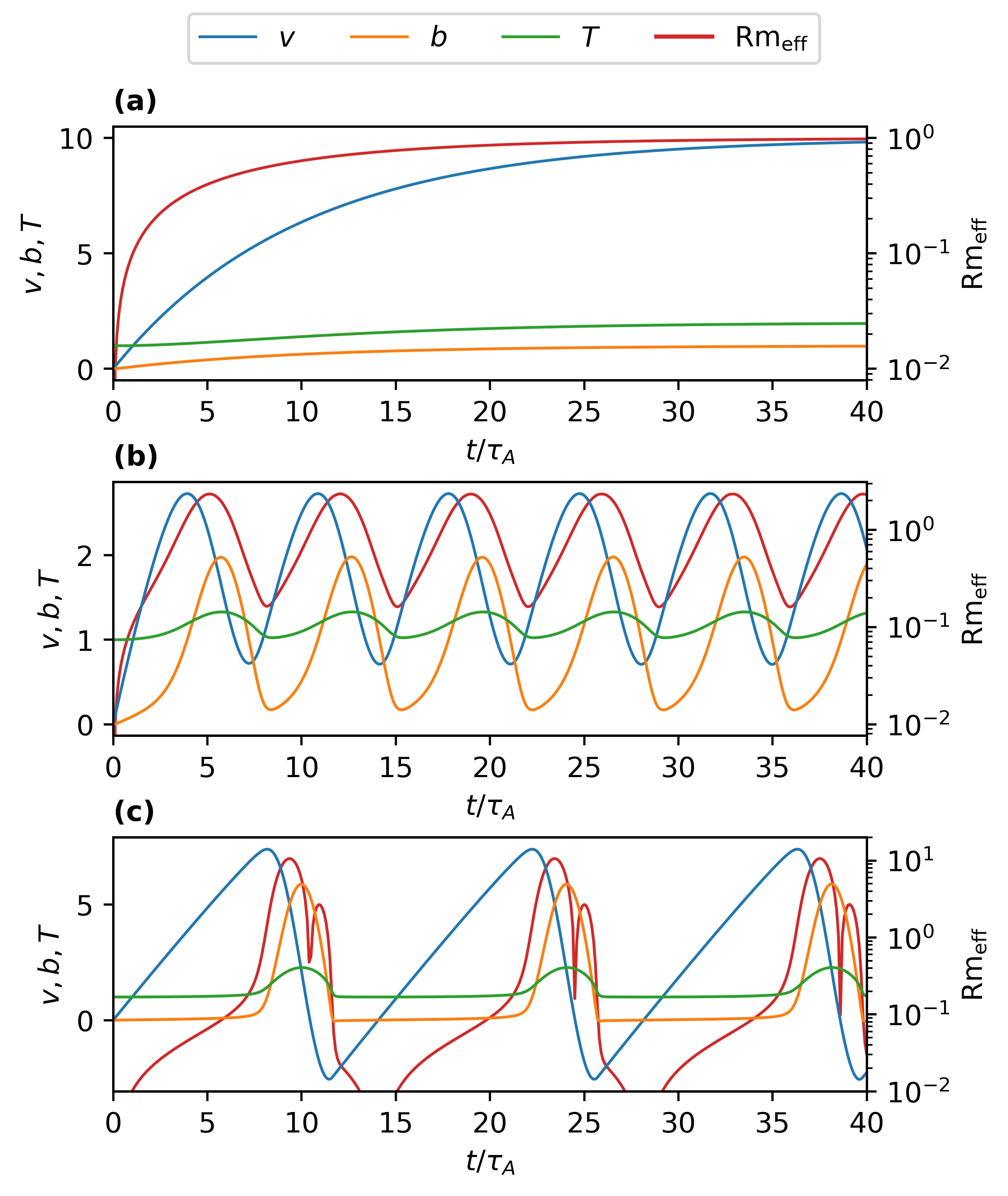}
    \caption{Time evolution of velocity $v$, toroidal field $b$, temperature $T$ and effective magnetic Reynolds number $\Rmeff = \Rm|v|/\etat$, showing the three different outcomes: (a) the system reaches steady-state ($\Rm$=0.1, $\tau$=1, $\Ec$=1, $\En$=0, $\dot{V}$=0.1); (b) oscillations ($\Rm$=0.1, $\tau$=1, $\Ec$=1, $\En$=10, $\dot{V}$=0.1) and (c) bursts ($\Rm$=0.01, $\tau$=10, $\Ec$=1, $\En$=10, $\dot{V}$=0.01).}
    \label{fig:time_series}
\end{figure}

Figure \ref{fig:time_series} shows three numerical integrations of equations~(\ref{eq:momentum})--(\ref{eq:temperature}) for different choices of parameters. In each case, we start the system from rest ($v=0$), with no toroidal magnetic field ($b=0$), and at the reference temperature ($T=1$). The top panel of Figure \ref{fig:time_series} shows an example in which the magnetic diffusivity does not depend on temperature ($E_\eta=0$). In this case, with low magnetic Reynolds number ($\Rm=0.1$), the system evolves smoothly to the steady state given by equation (\ref{eq:ss}).

The middle panel of Figure \ref{fig:time_series} shows the dramatically different behaviour that occurs when the magnetic diffusivity is temperature-sensitive. This integration has the same parameters as the top panel except that we now set $E_\eta =10$. Instead of a steady-state solution, we find sustained Alfvénic oscillations. While not a pure Alfvén wave due to the temperature-dependent Ohmic dissipation term, the solution shows the signature $\pi/2$ phase lag between $b$ and $v$ expected for Alfvén waves. 

It is important to note that during the oscillation, the effective Reynolds number $\Rmeff=\Rm|v|/\etat$ reaches values above $1$, even though the input parameter $\Rm<1$ (see red curve in Figure~\ref{fig:time_series}). In other words, the temperature increase is enough to reduce the magnetic diffusivity to a level where $\Rmeff$ becomes larger than unity. This temporary crossing of the effective magnetic Reynolds number from $\Rmeff<1$ to $\Rmeff>1$ seems to be a requirement for sustained oscillations to occur. We find that oscillations do not occur for other choices of parameters where the system either never reaches $\Rmeff$ values above unity, or has $\Rmeff$ values that stay above unity. In these cases, the system either evolves smoothly to steady state ($\Rmeff<1$) or shows damped oscillations to a steady-state ($\Rmeff>1$), similar to the situation with $E_\eta=0$.

    \begin{figure*}[t]
        \centering
        \includegraphics[width=0.9\linewidth]{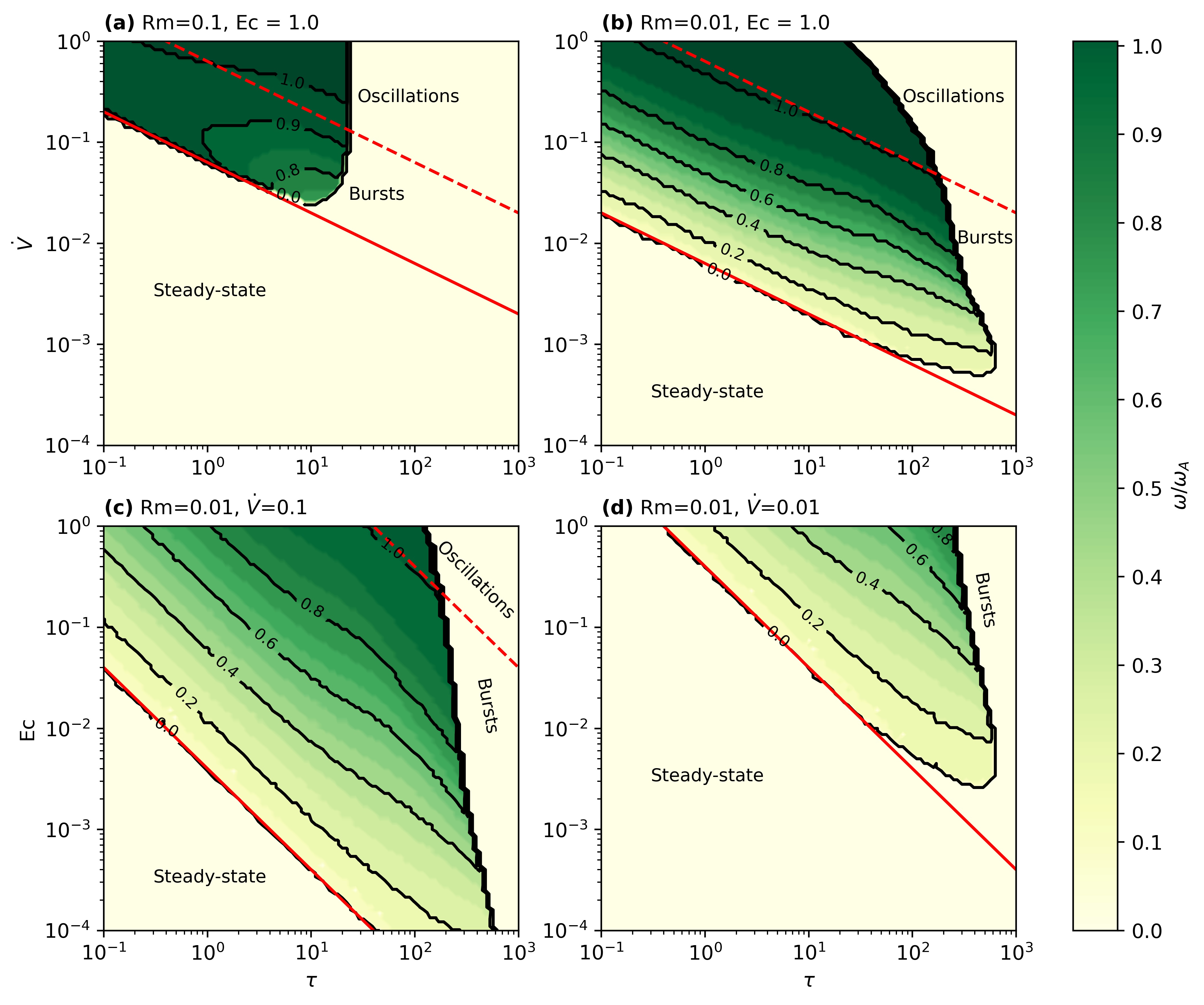}
        \caption{Angular frequency $\omega/\omega_A$ of the oscillation cycle as a function of $\dot{V}$ and $\tau$ for (a) $\Rm=0.1$, $\Ec=1.0$ and (b) $\Rm=0.01$, $\Ec=1.0$ and as a function of $\Ec$ and $\tau$ for (c) $\Rm=0.01$, $\dot{V}=0.1$ and (d) $\Rm=0.01$, $\dot{V}=0.01$. The magnetic diffusivity sensitivity is kept constant for all panels at $\En=10.0$. The black lines represent the isocontours of the angular frequency $\omega/\omega_A$. The solid red lines represent the instability criterion of equation~(\ref{eq:instability_boundary}); the system should be oscillating above it. The dashed red lines represent the analytic estimate of equation~(\ref{eq:burst_alfvenic}) for the boundary between bursts and Alfvénic oscillations.}
        \label{fig:contours}
    \end{figure*}

The bottom panel of Figure \ref{fig:time_series} shows a different behaviour in which the time-dependence takes the form of regular short-duration bursts. This case is intermediate between the solutions with Alfvénic oscillations and those that evolve smoothly to a steady-state. Compared to the previous example, this case has a longer cooling timescale, slower forcing, and lower $\Rm$. The long phase between bursts, lasting for more than 10 Alfvén times, can be understood as a quasi-steady-state in which the system slowly evolves along a sequence of steady-state solutions satisfying equation (\ref{eq:ss}) with increasing temperature as the system slowly heats up. Eventually, the temperature becomes large enough that the TRI occurs, driving the system to $\Rmeff>1$. 
The magnetic field then undergoes a rapid growth as it is now coupled to the flow, the resulting large magnetic torque then leads to a rapid drop in the velocity, an Alfvén oscillation follows, and eventually the cycle restarts. \review{In the example shown, the system oscillates for only half a period after the onset of the burst, but in other cases multiple damped oscillations can occur.} Depending on the parameters, we find that the interval between bursts varies smoothly from large to smaller values comparable to the Alfvén period, at which point the bursts smoothly transition to Alfvénic oscillations.

\subsection{Domains of instability and analytic estimates} \label{sec:regimes_domains}

Panels (a)-(b) of Figure~\ref{fig:contours} show the frequency of the oscillation cycle as a function of the forcing strength $\dot{V}$ and cooling time $\tau$, for two different values of $\Rm$. The other parameters are fixed at $E_\eta=10$ and $\Ec=1$. The instability occurs in the region with $\omega>0$ (green shading), at small $\tau$ and large $\dot V$. In the region where $\omega=0$ (pale yellow), the solution goes to steady-state and sustained oscillations do not occur.

Since the sustained oscillations require the effective magnetic Reynolds number to cross above and below one, we require $\Rm<1$ for instability. Indeed, in Figure~\ref{fig:contours}, we see that the instability region shrinks in size as $\Rm$ increases. It is noteworthy that we see an extended region of bursts (with $\omega<\omega_A$) only for the lower $\Rm$ case; at higher $\Rm$, the frequencies are close to $\omega_A$ throughout the unstable region. As a different slice of parameter space, panels (c)-(d) of Figure \ref{fig:contours} show the frequency in the $\Ec$--$\tau$ plane, this time with fixed $\Rm=0.01$ and $\dot{V}=0.1$. This shows that there exists a considerable region parameter space within which we find sustained oscillations, even when $\Ec$ takes values $\ll 1$.

We can understand the location of the unstable region with simple analytic arguments. First, in the long build-up phase of the oscillation, $\Rm \ll 1$ and magnetic drag is weak so that the velocity increases linearly with time,  $v\approx\dot{V}t$. Assuming steady-state in equation~(\ref{eq:induction}), the corresponding magnetic field is $b\approx \Rm \dot{V} t / \etat$. We can thus rewrite equation~(\ref{eq:temperature}) as
\begin{equation}\label{eq:temperature2}
\frac{dT}{dt} = -\frac{(T-1)}{\tau} + \frac{\Ec \Rm^2 (\dot{V}t)^2}{\tilde{\eta}}
\end{equation}
during the build-up phase. Direct numerical integration of equation (\ref{eq:temperature2}) shows good agreement with our integrations of the full equations until the moment when the TRI begins.

The recurrence timescale is set by the time taken to reach the TRI. Making a linear perturbation $T\rightarrow T+\delta T$ in equation (\ref{eq:temperature2}) gives an expression for the recurrence time as the time at which $d\delta T/dt$ transitions from being negative to positive (i.e. when $d\delta T/dt=0$),
\begin{equation}\label{eq:recurrence_timescale}
\tau_\mathrm{rec}^2 = \frac{\etat T^2}{\dot{V}^2 \tau \Rm^2 \Ec \En}.
\end{equation}
Comparing this timescale to the Alfvén timescale, $\tau_A=1/\Rm$, the transition from bursts to Alfvénic oscillations occurs when
\begin{equation}\label{eq:burst_alfvenic}
\dot{V}^2 \tau = 4\frac{\etat T^2}{\Ec \En},
\end{equation}
plotted as the dashed line in Figure \ref{fig:contours} (we have assumed small temperature changes, $T\approx 1$ and $\etat\approx 1$, when plotting the dashed line and have inserted a prefactor of 4 to better agree with the numerical results in this limit).

Figure \ref{fig:contours} shows that the system becomes stable at low values of $\dot{V}$. This instability boundary corresponds well with the location where the recurrence time $\tau_\mathrm{rec}$ starts to become longer than the time the system takes to reach the steady-state described by equation~(\ref{eq:ss}), $\tau_{ss}$. Since $v\approx \dot{V}t$ in the build-up phase, we find that $\tau_{ss}\approx \etat/\Rm^2$. Requiring $\tau_{ss}>\tau_\mathrm{rec}$ gives the criterion for instability as %
\begin{equation}\label{eq:instability_boundary}
\dot{V}^2 \tau > \frac{\Rm^2 T^2}{\etat \En \Ec}.
\end{equation}
We find that in our numerical solutions, the factor $T^2/\etat\approx 4$. The solid line in Figure \ref{fig:contours} shows this relation assuming that $T^2/\etat=4$, and gives a good match to the instability boundary.

As well as stabilizing at low $\dot{V}$, we also find that the oscillations stabilize at large $\tau$. Very approximately, values of the cooling time $\tau \gtrsim \tau_A = 1/\Rm$ stabilize the oscillations, since the temperature then does not change sufficiently during the oscillation period to enable the effective magnetic Reynolds number to alternate between values larger and smaller than unity.

\review{The value of $\En$ is kept constant throughout the panels of Figure \ref{fig:contours}. The criteria presented in equations (\ref{eq:burst_alfvenic}) and (\ref{eq:instability_boundary}) show that the effects of $\En$ and $\Ec$ on the system are degenerate as they always appear in pairs in the equations, and so our results can be rescaled to a different value of $\En$ as needed.}

Finally, for comparison we also derive the instability criterion from \cite{Menou2012instability} in our notation. The equivalent to our equation (\ref{eq:temperature2}) for the instability of \cite{Menou2012instability} is
\begin{equation}\label{eq:temperature_menou}
\frac{dT}{dt} = -\frac{(T-1)}{\tau} + \frac{\Ec \Rm^2 V^2}{\tilde{\eta}}
\end{equation}
where we have taken the weak drag limit with the wind velocity $V$ specified as a parameter and $b = \Rm\, V/\etat$. In this case, the instability criterion ($d\delta T/dt>0$) is
\begin{equation}\label{eq:menou}
    V^2 > \frac{\etat T^2}{\Rm^2\En \Ec\,\tau}.
\end{equation}
This is equivalent to equation (\ref{eq:instability_boundary}) with $V = \dot{V}\etat/ \Rm^2$, i.e.~the steady-state velocity under forcing $\dot{V}$.

\begin{figure*}[t]
    \centering
    \includegraphics[width=1.0\linewidth]{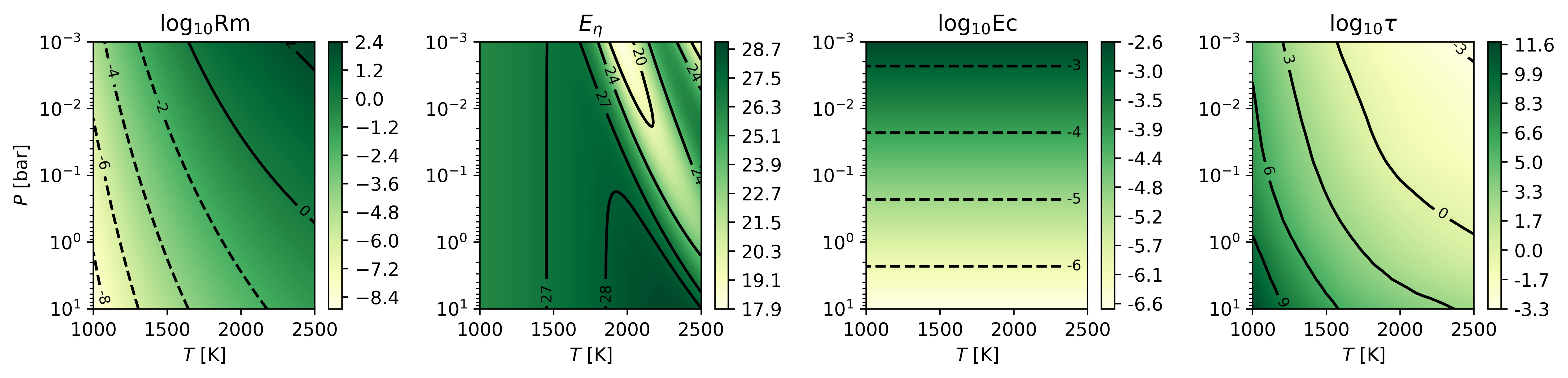}
    \caption{Values of $\Rm$, $\En$, $\Ec$, and $\tau$ shown as a function of temperatures and pressures relevant for HJ atmospheres. The magnetic Reynolds number is taken from equation~(\ref{eq:Rm_dimless_eq}). The temperature sensitivity of the magnetic diffusivity is defined as $\En \equiv -T_3 d\ln\eta/d\ln T$. The Eckert number is taken from equation~(\ref{eq:Ec_dimless_eq}). The radiative cooling time is taken from Table~2 of \citet{Showman2008} with a linear interpolation at higher temperatures and was turned into dimensionless form with the Ohmic time $t_{\rm Ohm}$ from equation~(\ref{eq:t_ohm}). We assume a surface gravity equal to that of Jupiter and a magnetic field of $B=10\ {\rm G}$ (note that $\Rm\propto B$, $\Ec\propto B^2$;  $\En$ and $\tau$ are independent of $B$). The black lines represent the isocontours of the presented parameters in each panel.}
    \label{fig:params_contours}
\end{figure*}

\section{Application to hot jupiters}\label{sec:application}

\subsection{Parameter values}

In the previous section we showed that instability occurs only for particular ranges of the parameters $\Ec$, $\En$, $\Rm$, $\tau$, and $\dot{V}$. We now estimate the expected values of these parameters in the atmosphere of a HJ. Values of $\Ec$, $\En$, $\Rm$, and $\tau$ are shown in Figure \ref{fig:params_contours} as a function of pressure and temperature. 

Figure \ref{fig:params_contours} shows that the temperature sensitivity of the magnetic diffusivity is indeed large, and lies within a narrow range of values. We define $E_\eta \equiv T_3 d\ln\eta/d\ln T$, where $T_3=T/1000\ {\rm K}$, and calculate the magnetic diffusivity $\eta=230 \sqrt{T/{\rm K}}/x_e\ {\rm cm^2\ s^{-1}}$ including contributions to the ionization fraction from species up to Ni (following \citealt{Menou2012instability}). We find $\En$ takes values between $\approx 18$ and $29$ for $10^{-3}<P<10\ {\rm bar}$ and $1000<T<2500\ {\rm K}$. Indeed for $T\lesssim 1800\ {\rm K}$, where the ionization fraction is dominated by potassium, $x_e\propto T^{3/4}\exp(-25.2/T_3)$ (e.g. \citealt{Perna2010b} equation~(1)), giving $E_\eta\approx 25$ almost independent of pressure.

The Eckert number $\Ec=v_A^2/c_PT$ is an important parameter for the instability because it determines whether Ohmic dissipation is able to heat the gas sufficiently quickly. The magnetic field strength of HJs is uncertain. We scale our results to $B=10\ {\rm G}$ in this section, although note that field strengths exceeding $\sim 100\ {\rm G}$ may arise in HJ atmospheres \citep{Reiners2010,Cauley2019,Yadav2017}.
The Alfvén speed is
$v_A = B/(4\pi \rho)^{1/2} \approx 570\ {\rm cm\ s^{-1}}\ (B/10\ {\rm G})\, T_3^{1/2} (P/1\,{\rm bar})^{-1/2}$. For the heat capacity, we assume pure molecular hydrogen with an ideal gas equation of state, giving $\rho = 2m_p P/k_BT \approx 2.42\times 10^{-5}\ {\rm erg\ cm^{-3}}\ T_3^{-1}\,(P/1\,{\rm bar})$ and $c_P\approx (7/2)k_B/2 m_p\approx 1.45\times 10^8\ {\rm erg\ K^{-1}\ g^{-1}}$. 
Therefore $\Ec\approx B^2/14\pi P$, or 
\begin{equation}\label{eq:Ec_dimless_eq}
\Ec\approx 2.3\times 10^{-6}\ \left(\frac{B}{10 \rm G}\right)^2 \left(\frac{P}{1\rm bar}\right)^{-1}.
\end{equation}
We see in Figure \ref{fig:params_contours} that $\Ec$ takes larger values at low pressure, suggesting that instability is more likely there, as also emphasized by \cite{Menou2012a} for the thermo-resistive instability. 

The magnetic Reynolds number $\Rm = v_AH/\eta$ can be written in terms of the ionization fraction as
\begin{equation}\label{eq:Rm_dimless_eq}
    \Rm \approx 0.1\,\left(\frac{x_e}{10^{-7}}\right)\ T_3 \left(\frac{B}{10 \rm G}\right)\,\left(\frac{P}{1 \rm bar}\right)^{-1/2},
\end{equation}
where we take the length scale $H$ to be the pressure scale height $H\approx k_BT/2m_pg\approx 1.6\times 10^7\ {\rm cm}\ T_3\,(g_J/g)$, normalized to the surface gravity of Jupiter $g_J=GM_J/R_J^2$. Figure \ref{fig:params_contours} shows that $\Rm$ can take on a large range of values since $x_e$ is so sensitive to the local temperature. For example at $T\approx 1200\ {\rm K}$, $x_e\sim 10^{-12}$--$10^{-10}$ for pressures in the range $10\ {\rm bar}$--$1\ {\rm mbar}$, while for $T\approx 2000\ {\rm K}$, $x_e\sim 10^{-8}$--$10^{-7}$ in the same pressure range.


For the radiative cooling timescale, we follow \cite{Menou2012a} and refer to the tabulated values in \cite{Showman2008} (Table 2 in that paper).
For $T\approx 1200\ {\rm K}$, the approximate cooling timescales given there are $\sim 3000\ {\rm s}$ ($\sim 10^6\ {\rm s}$) for a pressure of $1\ {\rm mbar}$ ($1\ {\rm bar}$). 
For $T\approx 2000\ {\rm K}$, these timescales become $\sim 300\ {\rm s}$ ($\sim 10^5\ {\rm s}$). Our dimensionless parameter $\tau$ is the ratio of the cooling timescale to the Ohmic diffusion time $t_{\rm Ohm}\approx H^2/\eta$, which we define using the pressure scale height as the lengthscale,
\begin{equation}\label{eq:t_ohm}
    t_{\rm Ohm}\approx 3.5\times 10^3\ {\rm s}\ \left(\frac{x_e}{10^{-7}}\right)\ T_3^{3/2} \left(\frac{g}{g_J}\right)^{-2}.
\end{equation}
Proportional to $x_e$, the Ohmic time spans a large range of values, ranging from $\sim 10^{-2}\ {\rm s}$ to hours depending on the local physical conditions.
Normalizing the radiative cooling time by the Ohmic time, we obtain $\tau$, shown in the last panel of Figure \ref{fig:params_contours}.

Perhaps the most uncertain parameter is the forcing $\dot{V}$ since this depends on the overall flow dynamics (and indeed the back reaction of the dynamics in the equatorial plane on the global flow). The units of $\dot{V}$ are $v_A/t_{\rm Ohm}$ which is

\begin{eqnarray}\label{eq:Vdotunits}
    \frac{v_A}{t_{\rm Ohm}} &\approx& 0.16\ {\rm cm\ s^{-2}}\ \left(\frac{B}{10 \rm G}\right)\, T_3^{-1} \left(\frac{P}{1 \rm bar}\right)^{-1/2}\nonumber\\
    & &\times \left(\frac{x_e}{10^{-7}}\right)^{-1}\left(\frac{g}{g_J}\right)^{2},
\end{eqnarray}
i.e., the ratio of the acceleration in cgs units to equation (\ref{eq:Vdotunits}) gives our parameter $\dot{V}$. Values of acceleration can be estimated from Figure~9 of \citet{Rogers2014b}, where the mean accelerations as a function of latitude for some of their models are shown. 
Taking a value of $5\times 10^{-8}\ {\rm g\ cm^{-2}\ s^{-2}}$ for the volumetric force from their Figure~9, and the corresponding pressure and temperature of $\approx 10\ {\rm mbar}$ and $\approx 1300\ {\rm K}$ to calculate the density, we obtain an acceleration $\approx 0.3\ {\rm cm\ s^{-2}}$ ($\approx 0.3\ {\rm km\ s^{-1}\ day^{-1}}$). For this value of acceleration, and taking the ionization fractions above, we find $\dot{V}\sim 10^{-6}$--$10^{-5}$ at $T\approx 1200\ {\rm K}$, and $\dot{V}\sim 0.01$ at $T\approx 2000\ {\rm K}$. Figure~2 of \citet{Rogers2017b} gives about a factor of ten smaller acceleration, showing velocity variations of $\sim 10^4\ {\rm cm\ s^{-2}}$ on timescales of $\sim 10^6\ {\rm s}$, suggesting accelerations of $0.01\ {\rm cm\ s^{-2}}$ at the mbar pressure level.

\subsection{Unstable regions as a function of pressure and temperature}

\begin{figure}[t]
    \centering
    \includegraphics[width=1.0\linewidth]{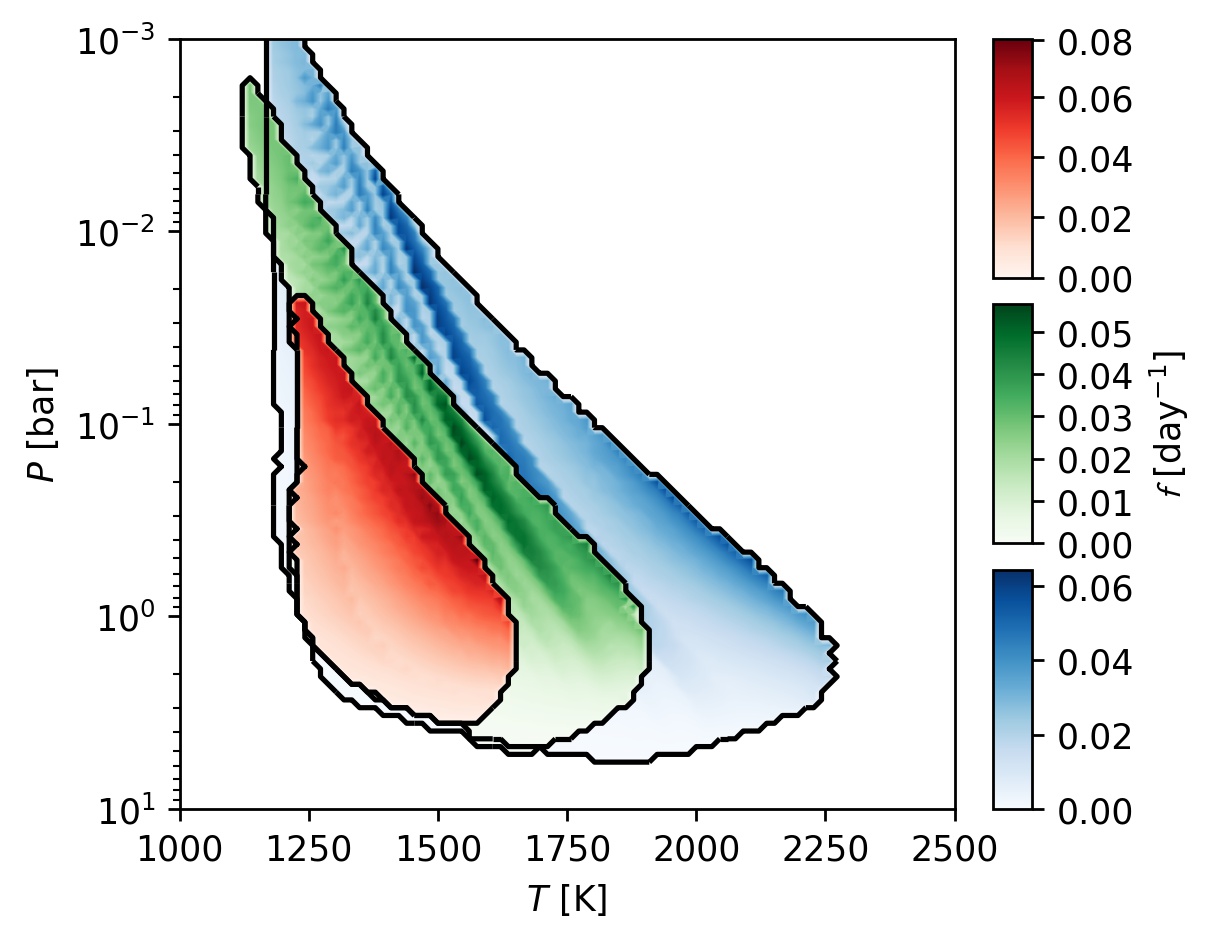}
    \caption{\review{Frequency $\bm f$ in day$^{-1}$} of the instability domain for the atmospheres of HJ with magnetic field strength of $B=10~\rm G$ and $B=40~\rm G$ and an acceleration of $\dot{v}=1~\rm cm~s^{-2}$ in green and red respectively. The blue region represents the results for $B=3~\rm G$ and $\dot{v}=1.4~\rm cm~s^{-2}$. The parameters used to solve equations (\ref{eq:momentum})--(\ref{eq:temperature}) were obtained the same way as in Figure \ref{fig:params_contours}. For $\dot{V}$, we applied the exponential decay, $\exp(-P/{\rm bar})$ and used the scaling presented in equation (\ref{eq:Vdotunits}) to transform $\dot{v}$ to its dimensionless counterpart, $\dot{V}$. \review{Frequencies are recast in dimensional units using the Alfv\'en time, as given by equation~(\ref{eq:alfven_time})}. The black lines represent the edges of the unstable regions.}
    \label{fig:criteria_P-T}
\end{figure}

We present in Figure \ref{fig:criteria_P-T} the \review{oscillation frequency} and instability regions for different choices of $B$ and $\dot v$. The red region has a magnetic field strength of $B = 40 \rm ~G$ and an acceleration of $\dot{v} = 1 \rm ~cm ~s^{-2}$, the green region has $B = 10 \rm ~G$ and also has $\dot{v} = 1 \rm ~cm ~s^{-2}$. The blue region has $B = 3 \rm ~G$ and $\dot{v} = 1.4 \rm ~cm ~s^{-2}$. The values of the magnetic fields are the same as in Figure 1 of \citet{Menou2012instability}, and the values of $\dot{v}$ have the same relative differences. The values of the necessary parameters are calculated as in Figure \ref{fig:params_contours} for each choice of $B$. For $\dot{v}$, we applied an exponential decay with respect to pressure, \review{multiplying the values of $\dot{v}$ above by $\exp(-P/{\rm bar})$,} and used equation (\ref{eq:Vdotunits}) to calculate the dimensionless parameter $\dot{V}$.

\begin{figure}[t]
    \centering
    \includegraphics[width=1.0\linewidth]{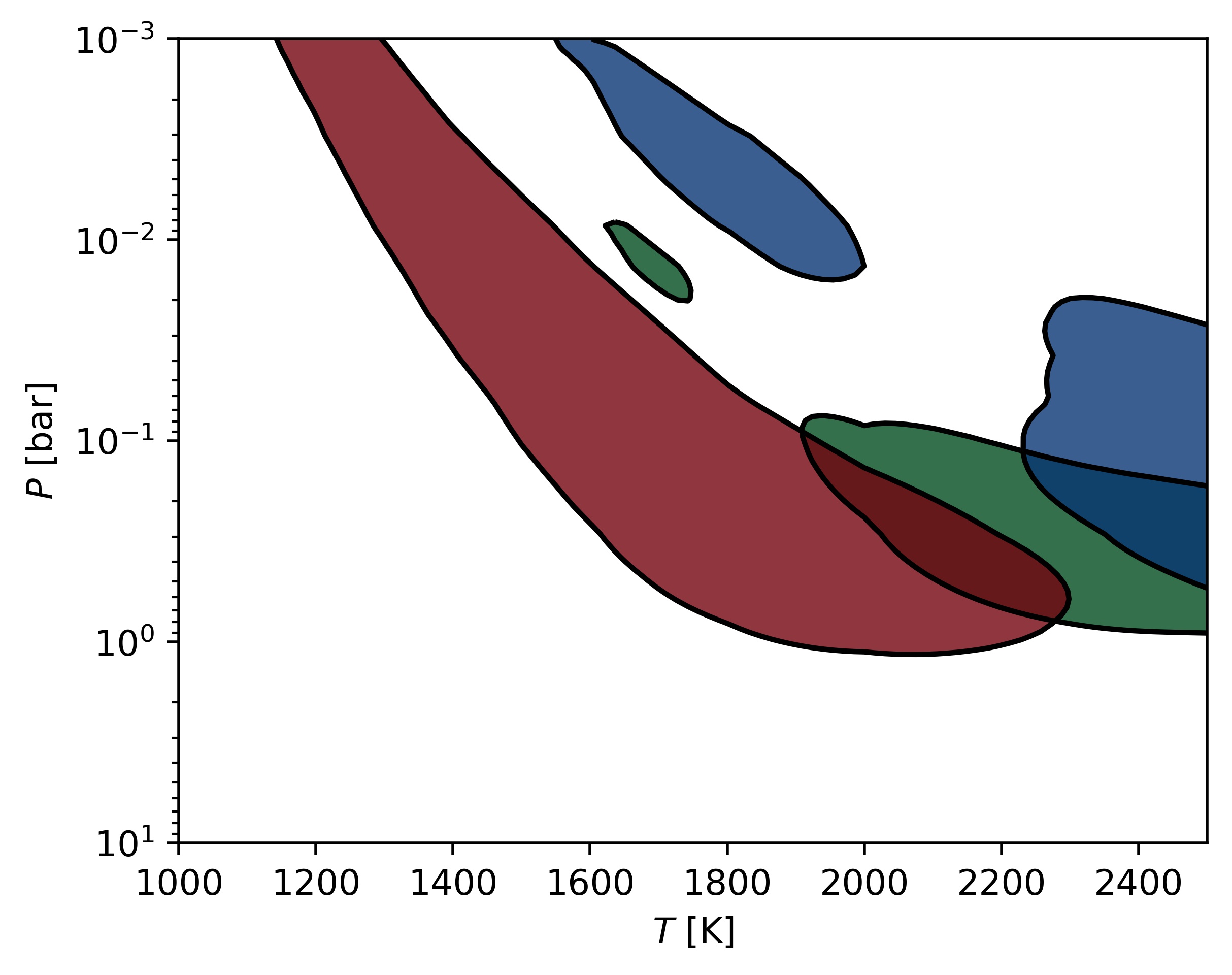}
    \caption{Thermo resistive instability domain using Equation (\ref{eq:menou}) with $\etat T^2=1$, combined with Equation (7) of \citet{Menou2012instability}. 
    The default magnetic drag law ($n_s = n_e = 1$) was used, as well as the exponential decay at higher pressures; $\exp(-P/{\rm bar})$. The green and red regions correspond to input zonal velocities of $V_\phi = 7 \rm ~km~s^{-1}$ with magnetic fields of $B=10$ and $40~\rm G$, respectively. The blue region represent the instability region for $V_\phi = 10 \rm ~km~s^{-1}$ and $B=3~\rm G$. These parameters were chosen to replicate Figure 1 of \citet{Menou2012instability} with equation (\ref{eq:menou}). The black lines represent the edges of the unstable regions.}
    \label{fig:criteria_P-T_menou}
\end{figure}


The instability regions are well defined within the mbar to a few bars pressure levels, and within 1250 to 2300~K. The unstable regions are shaded according to the frequency. We note that none of the simulations for these parameters reach sustained Alfvénic oscillations, and are all within the burst regime. \review{The bursts usually take the form of a series of decaying Alfvén waves after reaching the critical point.} Only the hottest part of the blue region comes close to the Alfvénic regime, where oscillations reach values slightly above $0.75 ~\omega/\omega_A$.

\review{The stripes that can be seen in Figure \ref{fig:criteria_P-T} are related to the number of Alfvén oscillations that occur following a burst. We find that the sequence of damped Alfvén oscillations terminates, and the growth of velocity due to $\dot{V}$ resumes, either at a maximum or minimum of velocity, which corresponds to the time in the oscillation when $b=0$ and ohmic heating vanishes. If the system reaches the growth phase when $v$ is at a minimum, then it needs to reach the critical velocity from negative values, like in panel (c) of Figure \ref{fig:time_series}. However, if the growth phase starts when $v$ is at a maximum and therefore positive, the system will have a head start in reaching the critical velocity, reducing the overall period and increasing the frequency of the TRI. As the parameters are varied, there are discrete transitions in the number of half-oscillations following the burst, leading to the jumps in frequency that can be seen in Figure \ref{fig:criteria_P-T}.}

\review{We have also plotted the instability regions corresponding to the analysis of \cite{Menou2012instability} (given by equation [\ref{eq:menou}])} for different input zonal velocities and magnetic field strengths in Figure \ref{fig:criteria_P-T_menou}. We note that we used different values for the radiative cooling timescales, shifting our results to higher temperatures compared to \citet{Menou2012instability}. However, our results still show the same behaviour; a higher magnetic field results in an instability region at lower temperatures (red), and a weaker field yields an instability at higher temperatures (blue). The green and blue instability domains are divided into two distinct regions. These separations would disappear if we were to relax our criterion by only $10\%$.

Both Figure \ref{fig:criteria_P-T} and \ref{fig:criteria_P-T_menou} show unstable regions. We have represented the analogous regions with the same colours. There are many resemblances to note. First, a stronger magnetic field will shift the TRI to lower temperatures as the interaction between flow and the field gets stronger with the latter. A larger acceleration $\dot{v}$, or zonal velocity $V_\phi$, makes the instability region larger. The overall shape the the regions are also similar. Indeed, they all resemble a titled droplet, meaning that our results and those of \citet{Menou2012instability} agree to some extent about the areas of instability surrounding a point in the $P-T$ space.

There are also striking differences between Figure \ref{fig:criteria_P-T} and \ref{fig:criteria_P-T_menou}. Arguably the most important disagreement is that with a larger field (red) our instability region is smaller than for a smaller field (green). This boils down to the difference in nature of the instabilities. The instability presented by \citet{Menou2012instability} and in Figure \ref{fig:criteria_P-T_menou} is considered activated when a thermal runaway occurs. The model presented in this paper is considered unstable when an oscillatory steady state is reached. In our model a strong magnetic field is less conducive  to producing oscillations, but is more inclined to produce a thermal runaway in the thermo-resistive framework of \citet{Menou2012instability}. Our dynamical model's unstable regions are also shifted towards higher pressures and lower temperatures. There is also much more overlap between the different coloured regions in Figure \ref{fig:criteria_P-T} than in Figure \ref{fig:criteria_P-T_menou}. While $\dot{v}$ and $V_\phi$ play analogous roles in the two models, pushing the system to the critical point, their fundamental natures are very contrasting.

\section{Discussion}

It is no surprise that a sheared fluid threaded by a magnetic field has an oscillatory response to steady forcing. However, for a temperature-independent magnetic diffusivity, these oscillations are transient: Ohmic losses damp the oscillations and the system evolves to a steady state in which the forced winding of the field lines is balanced by magnetic diffusion (e.g.~for a similar situation in stellar spin down see \citealt{Charbonneau1993}). We find that a temperature-dependent magnetic diffusivity introduces new solutions. The local model introduced in \S\ref{sec:model} shows that the interplay of Ohmic diffusion and the temperature-dependent diffusivity leads to long-lived sustained oscillations in which the magnetic Reynolds number alternates between large and small values during an oscillation cycle (Figure ~\ref{fig:time_series}).

This dynamical thermo-resistive instability may be relevant for some regions of HJ atmospheres (Figure ~\ref{fig:criteria_P-T}). We find that sustained oscillations occur when (1) the magnetic Reynolds number $\Rm<1$, so that there is opportunity to generate an induced field and Ohmic heating; (2) the cooling time is short compared to the Alfvén timescale, so that the gas temperature can respond to Ohmic heating on the timescale of an Alfvén oscillation, and (3) for values of $\Ec=v_A^2/c_PT$ approaching unity, so that Ohmic heating is able to raise the temperature efficiently. These requirements lead to unstable regions at intermediate temperatures $T\approx 1500\ {\rm K}$ and lower pressures $P\lesssim 0.1\ {\rm bars}$, for lower magnetic field strengths $B\lesssim 10\ {\rm G}$. 

The timescale for the oscillations is set by the Alfvén time across a pressure scale height $t_A=H/v_A$, which is 
\begin{equation}\label{eq:alfven_time}
    t_A \approx 10.5\ {\rm h}\ T_3^{1/2} \left(\frac{B}{10 \rm G}\right)^{-1}\,\left(\frac{P}{1 \rm bar}\right)^{1/2}\left(\frac{g}{g_J}\right)^{-1}.
\end{equation}
The sustained oscillations can take the form of modified Alfvén oscillations with period $\approx t_A$ (e.g.~middle panel of Figure ~\ref{fig:time_series}), or bursts of oscillations separated by long intervals of duration $\gg t_A$ (e.g.~lower panel of Figure~\ref{fig:time_series}), which could lead to interesting observable effects. \review{For the parameters expected in HJ atmospheres, we find that the solutions are in the burst regime. There are two competing effects that determine the oscillation frequency. One is that the Alfvén time is shorter at lower pressures, as described by equation (\ref{eq:alfven_time}). The other is that the oscillations become closer to the Alfvenic regime at higher pressure, which shortens the oscillation period at higher pressures. The resulting oscillation frequencies are shown in Figure \ref{fig:criteria_P-T} for different cases.}
Were such oscillations observed, it would be possible to use our results to put constraints on the strength of the magnetic field of the planet observed, knowing the depth and temperature of the observed region, as well as the period of the oscillation. Possible degeneracies may arise from such an analysis, as Figure \ref{fig:criteria_P-T} suggests, therefore our model may only be able to impose an upper limit on the field's strength.

Our work generalizes the analysis of \cite{Menou2012instability} to include the dynamical effect of the magnetic field. Whereas \cite{Menou2012instability} assumed a given background wind velocity, we have instead assumed a given background acceleration that drives the flow. This is motivated by the behaviour seen in global dynamical models, which show net eastward or westward accelerations in the equatorial region (eastward in hydrodynamic models, westward in magnetized cases; \citealt{Hindle2021b}). The resulting instability criterion (equation ~[\ref{eq:instability_boundary}]) is equivalent to the instability criterion of \cite{Menou2012instability} with the wind velocity taken to be the steady-state velocity under the applied forcing (equation~[\ref{eq:ss}]). Putting in the units, this is a velocity $V = \dot{V} \eta/v_A^2=\dot{V}t_{\rm drag}$, i.e. the velocity acquired in a time
\begin{equation}\label{eq:tdrag}
t_\mathrm{drag}\approx 2.5\ {\rm d} \ \left(\frac{x_e}{10^{-7}}\right)^{-1}  T_3^{-1/2}\left(\frac{P}{1 \rm bar}\right)\left(\frac{B}{10 \rm G}\right)^{-2},
\end{equation}
is the magnetic drag timescale (see equation~[1] of \citealt{Rauscher2013}). This helps to understand why we prefer lower values of $B$: to achieve the $\sim {\rm km/s}$ velocities that \cite{Menou2012instability} requires for instability with an acceleration $\sim 1\ {\rm cm\ s^{-2}}$ requires a drag timescale of $\sim 10^6\ {\rm s}$. With a constant velocity rather than acceleration, \cite{Menou2012instability} found that increasing $B$ favoured instability; we find the opposite here (compare equations [\ref{eq:instability_boundary}] and [\ref{eq:menou}] and Figures \ref{fig:criteria_P-T} and \ref{fig:criteria_P-T_menou}).

Our results add support to the idea that variability should be a feature of magnetized HJ atmospheres, in particular at intermediate temperatures $T\approx 1500\ {\rm K}$. 
\review{Even the highest frequency simulations represent a period of oscillation of over 10 days. Such timescales are longer than typical integration times for observations and should therefore be resolved in observations of HJ atmospheres. However, the timescale of variability could be much shorter at lower pressures or higher magnetic field strengths (equation~(\ref{eq:alfven_time})). In particular, the decaying Alfvén waves following bursts could be challenging to resolve as the Alfvén time is smaller than the period between bursts.}

Spatially-extended MHD simulations of atmospheric dynamics including a temperature-dependent $\eta$ are needed to go beyond our local model and address whether the TRI plays a role in generating observable variability. In particular, the forcing acceleration $\dot{V}$ is put in by hand in our local model, but arises as part of the global atmospheric dynamics. Indeed, as explained in detail by \cite{Hindle2021b}, global Lorentz forces can lead to a reversal of the mean flow at the equator in magnetized atmospheres, i.e. changing the sign of $\dot{V}$ compared to hydrodynamic models. Similarly, our local analysis shows that the TRI is unstable only within a range of pressures; understanding the interaction of the low pressure unstable region with the deeper stable regions requires going beyond a local analysis. 

It is important to note that previous work on MHD models of HJ atmospheres has shown that variable flows arise even with a time-independent $\eta$ profile. \cite{Rogers2014b} found that as the influence of the magnetic field increased (either because of a hotter atmosphere or stronger magnetic field), the MHD effects changed from a drag force, to oscillatory mean flows, to a reversed (westward) mean flow. \cite{Hindle2021b} explain the reversal of the mean flow in terms of Lorentz forces associated with the toroidal magnetic fields generated by zonal flows. In addition, \cite{Rogers2017a} showed that the spatial gradient of $\eta$ can lead to magnetic field generation in an atmospheric dynamo. The temperature-dependence of $\eta$ is a further important ingredient to include in future MHD models of HJ atmospheric dynamics.\newline


\review{We thank the anonymous reviewer for helpful comments that improved this manuscript.} This work was supported by NSERC Discovery grants. RH, AC, and PC are members of the Centre de Recherche en Astrophysique du Québec (CRAQ).


\bibliography{mybib}{}
\bibliographystyle{aasjournal}

\end{document}